\documentclass[aps,pra,onecolumn,showpacs,amsmath,amssymb,floatfix,footinbib,superscriptaddress]{revtex4-1}
\usepackage{graphicx,natbib}
\usepackage{dcolumn}
\usepackage{bm}
\usepackage{amsmath,graphics,amssymb,psfrag}
\usepackage[colorlinks=true,linkcolor=blue,urlcolor=blue,citecolor=blue]{hyperref}
\usepackage[left]{lineno}
\usepackage{tikz,xcolor,hyperref}
\definecolor{lime}{HTML}{A6CE39}
\DeclareRobustCommand{\orcidicon}{%
    \begin{tikzpicture}
    \draw[lime, fill=lime] (0,0)
    circle [radius=0.16]
    node[white] {{\fontfamily{qag}\selectfont \tiny ID}};\draw[white, fill=white] (-0.0625,0.095)
    circle [radius=0.007];
    \end{tikzpicture}
    \hspace{-2mm}}
\foreach \x in {A, ..., Z}
{\expandafter\xdef\csname orcid\x\endcsname{\noexpand\href{https://orcid.org/\csname orcidauthor\x\endcsname}{\noexpand\orcidicon}}}

\begin{document}

\title{High quantum yields generated by a multi-band quantum dot photocell }
\author{Shun-Cai Zhao\orcidA{}}
\email[Corresponding author: ]{zsczhao@hotmail.com}
\affiliation{Department of Physics, Faculty of Science, Kunming University of Science and Technology, Kunming, 650500, PR China}

\author{Qi-Xuan Wu }
\affiliation{College English department, Faculty of foreign languages and culture, Kunming University of Science and Technology, Kunming, 650500, PR China}


\begin{abstract}
We perform the quantum yields in a  multi-band quantum dot (QD) photocell via doping an intermediate band (IB) between the conduction band (CB) and valence band (VB). Under two different sub-band gap layouts, the output power has a prominent enhancement than the single-band gap photocell and the achieved peak photo-to-charge efficiency reaches to 74.9\(\%\) as compared to the limit efficiency of 63.2\(\%\) via the IB approach in the theoretical solar cell prototype. The achieved quantum yields reveal the potential to improve efficiency by some effective theoretical approaches in the QD-IB photocell.
\begin{description}
\item[PACS numbers]42.50.Gy; 42.50.Ct
\item[Keywords]Quantum yields; photo-to-charge efficiency; multi-band quantum dot photocell
\end{description}
\end{abstract}
\maketitle
\section{Introduction}

Recently, the quantum photovoltaic system has attracted extensive attention because it is believed that its yield can be enhanced via some quantum mechanism\cite{1,2,3,4,5,6,7,8,9,10,11}. In the seminal work\cite{1}, it shows that the photocell power can achieve a substantial enhancement via the coherence induced by the driving field, even if the actual photo-to-charge efficiency of the photocell was not calculated. However, another mechanism of coherence without an external source, the noise-induced coherence or Fano-induced coherence can yield the enhancement of power in recent papers\cite{2,3}, in which the detailed balance\cite{12} was broken and the photocell power delivered to the load was increased.

As it is well known, two major factors limit the quantum yield of the solar cell prototype. One is losing excess kinetic energy of photo-generated carriers as multiple phonon emission due to the absorption of photons with the energy greater than the energy gap between the conduction and valence bands\cite{13}. The second one is the loss of photons with energy lower than the band gap. These result in the Shockley-Quiesser detailed balance limit and the energy conversion efficiency is found\cite{12} to be about 33\(\%\). Due to their quantum confinement in three dimensions resulting in the tunable size and accordingly tunable band gap, QDs are playing a vital role in the solar cells domain and have achieved a considerable attention in the recent decades\cite{14,15,16,17,18,19}. Not only that, but QDs have the great potential to significantly increase the photo-to-charge efficiency via the formation of IBs in the band gap of the background semiconductor, which can absorb photons with energy below the semiconductor band gap through transitions from the VB to the IB and from the IB to the CB \cite{20,21,22,23,24,25,26,27,28,29}. In the QD-IB solar cell approach, the limiting efficiency of 47\(\% \) at 1 sun, 63.2\(\% \) at full solar concentration\cite{23} were achieved, which exceeds the previous maximum 33\(\%\) efficiency proposed by Shockley and Queisser in 1961\cite{12}.

In this paper, we have theoretically calculated the quantum yields enhancement of the QD-IB photocell with an intermediate level. This QD-IB photocell can be introduced by the doping technology, and its prototype comes from the seminal quantum photovoltaic system\cite{1}. It is proven that the output power has a prominent enhancement in this QD-IB photocell as compared to the single gap QD photocell\cite{1,2,3,4,5}. And the quantum peak photo-to-charge efficiency comes to 74.9\(\% \), which also exceeding 63.2\(\% \) mentioned in Ref.\cite{23} is confirmed.

In order to clarify our work, this paper is organized as follows: we describe the quantum dot photocell model with an IB and deduce the power and photo-to-charge efficiency generated by this photocell model in Sec.2. And the power and photo-to-charge efficiency are evaluate by two different sub-band gap layouts with different parameters in Sec.3. Sec.4 our conclusions and outlook are presented.

\section{Quantum dot photocell model with an intermediate band}
The proposed QD-IB photocell model can be generated by the doping IB \(|c_{3}\rangle\) via the doping technology in the theoretical prototype mentioned in Ref.\cite{1}. Therefore, the band gap between the CB and the VB was divided into two sub-bands, i.e., the upper sub-band gap \(E_{i3(i=1,2)}\) and the lower sub-band gap \(E_{3b}\). And the doublet CBs \(|c_{1}\rangle\) and \(|c_{2}\rangle\) may be produced by splitting of two degenerate QD levels due to electron tunneling between two adjacent dots [see Fig. 1(a)]. The subdivided energy gaps can absorb the solar photons with energy below the semiconductor band gap shown in Fig. 1(a). Then, due to the IB \(|c_{3}\rangle\), photons with energy below the band gap can be absorbed to excite electrons from the VB to the IB and from the IB to CB. Fig. 1(b) is the corresponding energy levels diagram for the QD photocell model, in which the two upper levels \(|c_{1}\rangle\), \(|c_{2}\rangle\) depict the CBs and level \(|b\rangle\) is on behalf of the VB, and the level  \(|c_{3}\rangle\) is interpreted as the IB. When the electronic system interacts with radiation and phonon thermal reservoirs, the thermal solar photons with frequency midway between \(\omega_{i3} (i=1,2)\) and \(\omega_{3b} \) are assumed to directing onto the cell. They drive \(|c_{1,2}\rangle\) \(\leftrightarrow\) \(|c_{3}\rangle\) and \(|c_{3}\rangle\)\( \leftrightarrow\) \(|b\rangle\) transitions and have average occupation number \(n_{1}\) = \([exp( \frac{E_{i3}}{k_{B}T_{s}})-1]^{-1}\) and \(n_{3}\) = \([exp( \frac{E_{3b}}{k_{B}T_{s}})-1]^{-1}\), where \(E_{i3} = (E_{13} + E_{23})/2\), \(T_{s}\) is the solar temperature.
The ambient thermal phonons at temperature \(T_{a}\) drive the low-energy transitions \(|c\rangle\) \(\leftrightarrow\) \(|c_{1}\rangle\), \(|c\rangle\) \(\leftrightarrow\) \(|c_{2}\rangle\), and \(|v\rangle\) \(\leftrightarrow\) \(|b\rangle\). Their corresponding phonon occupation numbers are \(n_{2}\) = \([exp( \frac{E_{12}}{k_{B}T_{a}})-1]^{-1}\) and \(n_{4}\) = \([exp( \frac{E_{vb}}{k_{B}T_{a}})-1]^{-1}\), where \(E_{12} = (E_{1c} + E_{2c})/2\).

\begin{figure}[htp]
\center
\includegraphics[width=0.44\columnwidth]{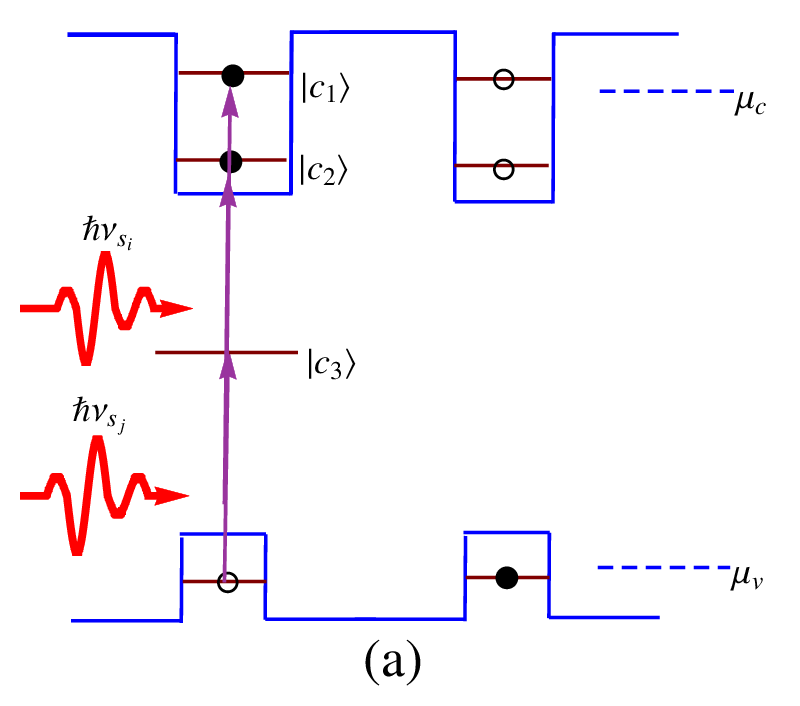 }\includegraphics[width=0.44\columnwidth]{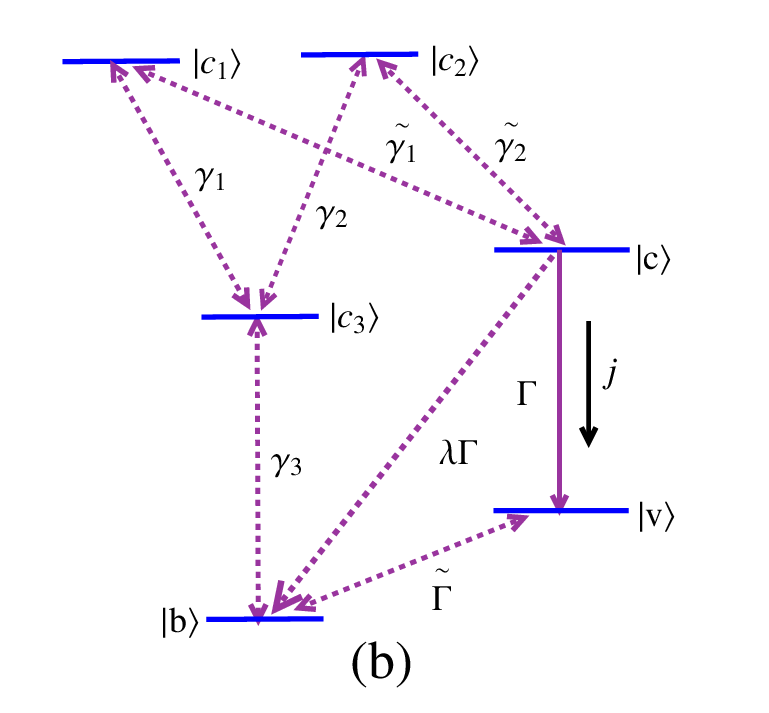 }
\caption{(color online) (a) Quantum dots having two upper level conduction band states, \(|c_{1}\rangle\)and  \(|c_{2}\rangle\).
The two different monochromatic solar photons(\(\nu_{s_{i}}\),  \(\nu_{s_{j}}\)) are tuned to the midpoint between the two upper levels and intermediate band \(|c_{3}\rangle\), the intermediate band \(|c_{3}\rangle\) and level \(|b\rangle\) in the valence band. The host semiconductor system, in which the
quantum dots are embedded, has effective Fermi energies \(\mu_{c}\) and \(\mu_{v}\) for the conduction and valence bands. (b) Corresponding
energy levels diagram of this QD-IB photocell model. Solar radiation drives transitions between the ground state \(|b\rangle\) and the intermediate band \(|c_{3}\rangle\), the intermediate band \(|c_{3}\rangle\) and two upper levels \(|c_{1}\rangle\), \(|c_{2}\rangle\). Transitions \(|c_{1}\rangle\), \(|c_{2}\rangle\) and  \(|v\rangle\), \(|v\rangle\) and \(|b\rangle\) are driven by ambient thermal phonons. Levels \(|c\rangle\) and \(v\rangle\) are connected to a load.}
\end{figure}
\label{Fig.1}

The interaction Hamiltonian in the rotating-wave approximation for the electronic system interacting with radiation and phonon thermal reservoirs is given as follows,

\begin{eqnarray}
&\hat{V}(t)=&\hbar \{\sum_{i} g_{i}[e^{i(\omega_{13}-\nu_{i})t}\hat{\sigma}_{13}+e^{i(\omega_{23}-\nu_{i})t}\hat{\sigma}_{23}]\hat{a}_{i}+ \sum_{j} g_{j} e^{i(\omega_{3b}-\nu_{j})t}\hat{\sigma}_{3b} \hat{a}_{j}+ \nonumber\\
       &&     \sum_{o}g_{o}[e^{i(\omega_{1c}-\nu_{o})t}\hat{\sigma}_{1c}+e^{i(\omega_{2c}-\nu_{o})t}\hat{\sigma}_{2c}]\hat{b}_{o} +\sum_{p}g_{p} e^{i(\omega_{bv}-\nu_{p})t}\hat{\sigma}_{bv} \hat{b}_{p}\}+h.c.
\end{eqnarray}

\noindent  where \(g_{i,j,o,p}\) are the coupling constants for the corresponding transition, \(\hbar\omega_{ij}=E_{i}-E_{j}\) is the energy spacing
between levels \(|i\rangle\) and \(|j\rangle\). \(\nu_{i}\) is the photon or phonon frequency, and \(\hat{a}_{i,j}\),  \(\hat{b}_{o, p}\) are the photon annihilation operators and phonon annihilation operators, respectively. \(\hat{\sigma}_{ij}(\hat{\sigma}_{ji})\) is the Pauli rise or fall operator. The equation of motion for the electronic density operator \(\rho\) reads

\begin{eqnarray}
&\dot{\rho}(t)=-\frac{i}{\hbar} Tr_{R}[\hat{V}(t),\rho(t_{0})\bigotimes\rho_{R}(t_{0}]-\frac{1}{\hbar^{2}}Tr_{R} \int^{t}_{t_{0}}[\hat{V}(t),[\hat{V}(t'),\rho(t')\bigotimes\rho_{R}(t_{0}]]dt',
\end{eqnarray}

\noindent where the density operator \(\rho_{R}\) describes the phonon thermal reservoirs. In the Weisskopf-Wigner approximation, we
assume that the levels \(|c_{1}\rangle\) and \(|c_{2}\rangle\) are degenerate and obtain the following master equations,
\begin{eqnarray}
&\dot{\rho_{11}}\!=\!&-\gamma_{1}[(1+n_{1})\rho_{11}-n_{1}\rho_{33}]-\tilde{\gamma}_{1}[(1+n_{2})\rho_{11}-n_{2}\rho_{33}]+\nonumber\\
                 &&\gamma_{12}(1+n_{1})Re[\rho_{12}]-\tilde{\gamma}_{12}(1+n_{2})Re[\rho_{22}],\nonumber\\
&\dot{\rho_{22}}\!=\!&-\gamma_{2}[(1+n_{12})\rho_{22}-n_{1}\rho_{33}]-\tilde{\gamma}_{2}[(1+n_{2})\rho_{22}-n_{2}\rho_{33}]+\nonumber\\
                  &&\gamma_{12}(1+n_{1})Re[\rho_{12}]-\tilde{\gamma}_{12}(1+n_{2})Re[\rho_{22}],\nonumber\\
&\dot{\rho_{12}}\!=\!&-\frac{1}{2}[(\gamma_{1}+\gamma_{2})(1+n_{1})+(\tilde{\gamma}_{1}+\tilde{\gamma}_{2})(1+n_{2})] \rho_{12}\nonumber\\
                 &&-\frac{1}{2}\gamma_{12}[(1+n_{1})\rho_{11}+(n_{1}+1)\rho_{22}-2n_{1}\rho_{33}]\nonumber\\
                 &&-\frac{1}{2}\tilde{\gamma}_{12}[(1+n_{2})\rho_{11}(1+n_{2})\rho_{22}-2n_{2}\rho_{33}]-\tau_{12}\rho_{12},\nonumber\\
&\dot{\rho_{33}}\!=\!&-\gamma_{3}[(1+n_{3})\rho_{33}-n_{3}\rho_{bb}]-(\gamma_{1}+\gamma_{2})n_{1}\rho_{33}+\gamma_{1}(1+n_{1})\rho_{11}\nonumber\\
                 &&+\gamma_{2}(1+n_{1})\rho_{22}+2\gamma_{12}(1+n_{1})Re[\rho_{12}], \\ \nonumber
&\dot{\rho_{cc}}\!=\!&\tilde{\gamma}_{1}[(1+n_{2})\rho_{11}-n_{2}\rho_{cc}]+\tilde{\gamma}_{2}[(1+n_{2})\rho_{22}-n_{2}\rho_{cc}]+2\tilde{\gamma}_{12}\nonumber\\ &&(1+n_{2})Re[\rho_{12}]-\Gamma(1+\lambda)\rho_{cc},\nonumber\\
&\dot{\rho_{vv}}\!=\!&\Gamma\rho_{cc}+\tilde{\Gamma}n_{4}\rho_{bb}-(1+n_{4})\rho_{cc}, \nonumber
\end{eqnarray}

\noindent where \(\gamma_{i}\) and  \(\tilde{\gamma}_{i}\) are modeled the electrons' decaying process as spontaneous decay rates of the corresponding transitions [see Fig.1(b)]. For maximum Fano coherence, \(\gamma_{12}=\sqrt{\gamma_{1}\gamma_{2}}\), \(\tilde{\gamma}_{12}\)=\(\sqrt{\tilde{\gamma}_{1}\tilde{\gamma}_{2}}\), while \(\gamma_{12}=\tilde{\gamma}_{12}\)=0 for no interference. In this model, levels \(|c\rangle\) and \(|v\rangle\) are connected to a load, and the load is yielding a decay of the level \(|c\rangle\) into level \(|v\rangle\) at a rate \(\Gamma\). The recombination between the acceptor and the donor is modeled by a decay rate of \(\lambda \Gamma\) [also see Fig. 1(b)], where \(\lambda\) is a dimensionless fraction, \(\tau_{12}\) is the decoherence time. And we focus on steady-state operation of the photocell model, voltage \(V\) across the load is expressed in terms of populations of the levels \(|c\rangle\) and \(|v\rangle\) as\cite{1,2,3}

\begin{eqnarray}
eV\!=\! E_{c}-E_{v}+k_{B}T_{a} \ln(\frac{\rho_{cc}}{\rho_{vv}}).
\end{eqnarray}

\noindent where \(T_{a}\) is the ambient temperature circumstance. The current \(j\) through the cell is interpreted as  \(j=e \Gamma\rho_{cc}\) and power delivered to the
load is P = \(j ¡¤ V \) which is supplied by the incident solar radiation power \(P_{s}\)\cite{1,2}, and their ratio describes the photo-to-charge efficiency of this QD-IB photocell.

\section{Power and photo-to-charge efficiency}

The photocell power enhanced by the noise-induced or Fano-induced coherence has been investigated detailedly by the former work\cite{2,6}. In the following, we abnegate the discussion on the output power influenced by the coherence, while turn to discuss the quantum power and photo-to-charge conversion efficiency under the condition with the maximum coherence interference,  $\gamma_{12}$= $\sqrt{\gamma_{1}\gamma_{2}}$, $\tilde{\gamma}_{12}$= $\sqrt{\tilde{\gamma}_{1}\tilde{\gamma}_{2}}$. And in present paper, the difference between this QD-IB photocell model and the QD photocell theoretical prototype in Ref.\cite{1} is the introduced IB. Therefore, the roles of different sub-band gaps in the quantum yield will attract much attention.

With the stationary solutions to the master equations(3), we explore the power and photo-to-charge efficiency through the
numerical calculations. And several typical parameters should be selected before the calculation. The energy gaps \(E_{1c}\)=\(E_{2c}\)=\(E_{vb}\)=\(0.005 ev\), \(E_{cv}\)=\(1.49 ev\), \(T_{a}\)=\(0.0259 ev\), \(T_{s}\)=\(0.5 ev\)\cite{1}. And other parameters are scaled by $\gamma$: $\gamma_{1}$=1.2$\gamma$, $\gamma_{2}$=0.8$\gamma$, $\gamma_{3}$=1.5$\gamma$, $\widetilde{\gamma}_{1}$=1.05$\gamma$, $\widetilde{\gamma}_{2}$=1.15$\gamma$, $\Gamma_{1}$=1.25$\gamma$, $\Gamma_{2}$=1.35$\gamma$.
The dimensionless fraction is set as \(\lambda\)=0.2. The Shockley and Queisser efficiency \cite{12} is accessible in semiconductors with band gaps ranging from about 1.25 ev to 1.45 ev. However, the solar spectrum contains photons with energies ranging from about 0.5 eV to 3.5 eV. Photons with energies ranging from about 0.5 ev to 1.25 ev is not absorbed and photons with energies within [1.45ev, 3.5ev] can incur the thermalization loss in the Shockley and Queisser efficiency. Therefore, the lower sub-band gap (\(E_{3b}\)) simulated as the solar spectrum ranging from 0.5 ev to 3.5 ev will be discussed in Fig.2, while the upper sub-band gap(\(E_{i3(i=1,2)}\)) between the CB and IB will select values from the range of [0.45 ev, 1.45 ev] with the output voltage being 1.5 volts in Fig.2 (And the data for Fig.2 are in the SI Word).

\begin{figure}[!t]
\centerline{\includegraphics[width=0.52\columnwidth]{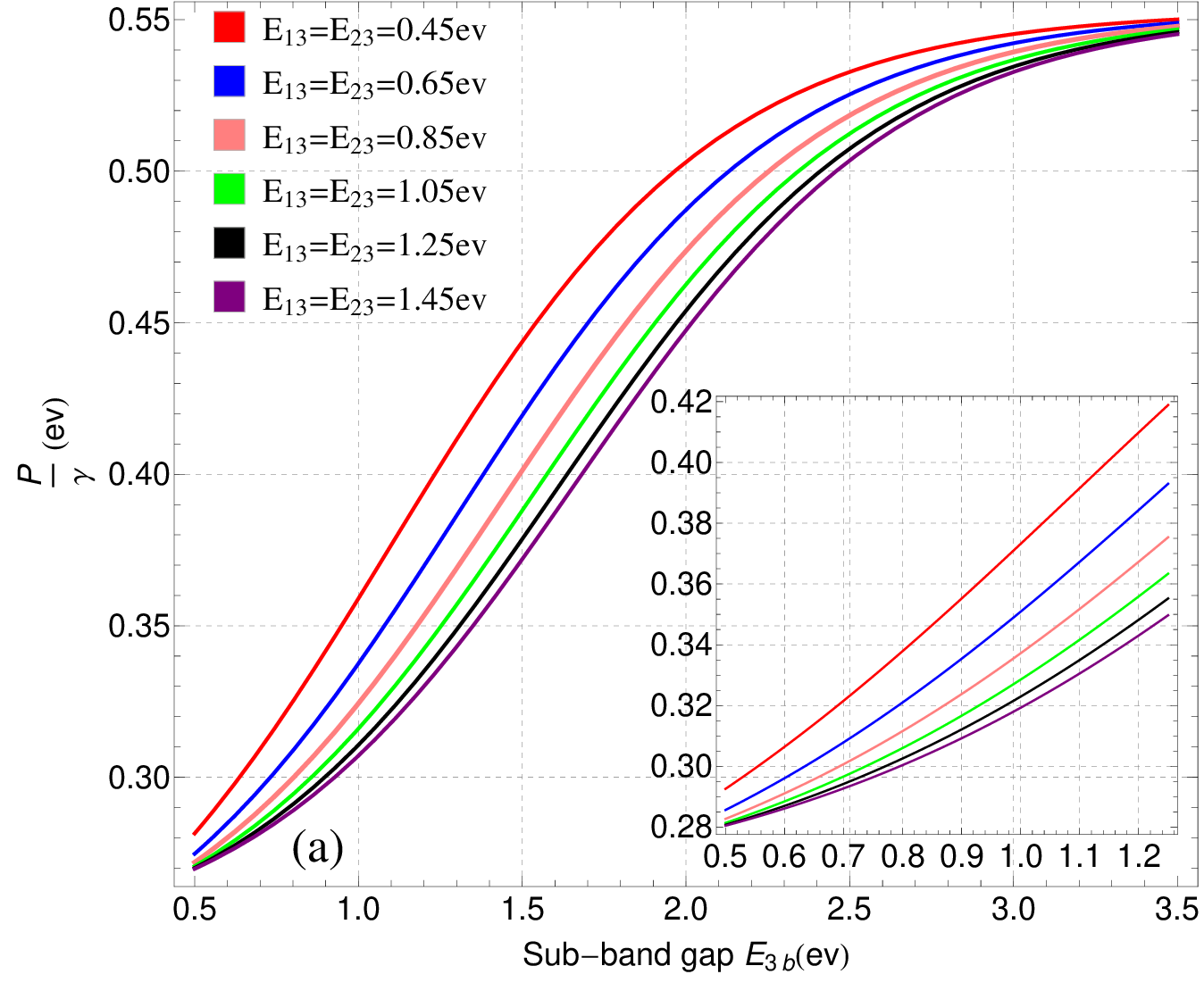 }~~~~\includegraphics[width=0.45\columnwidth]{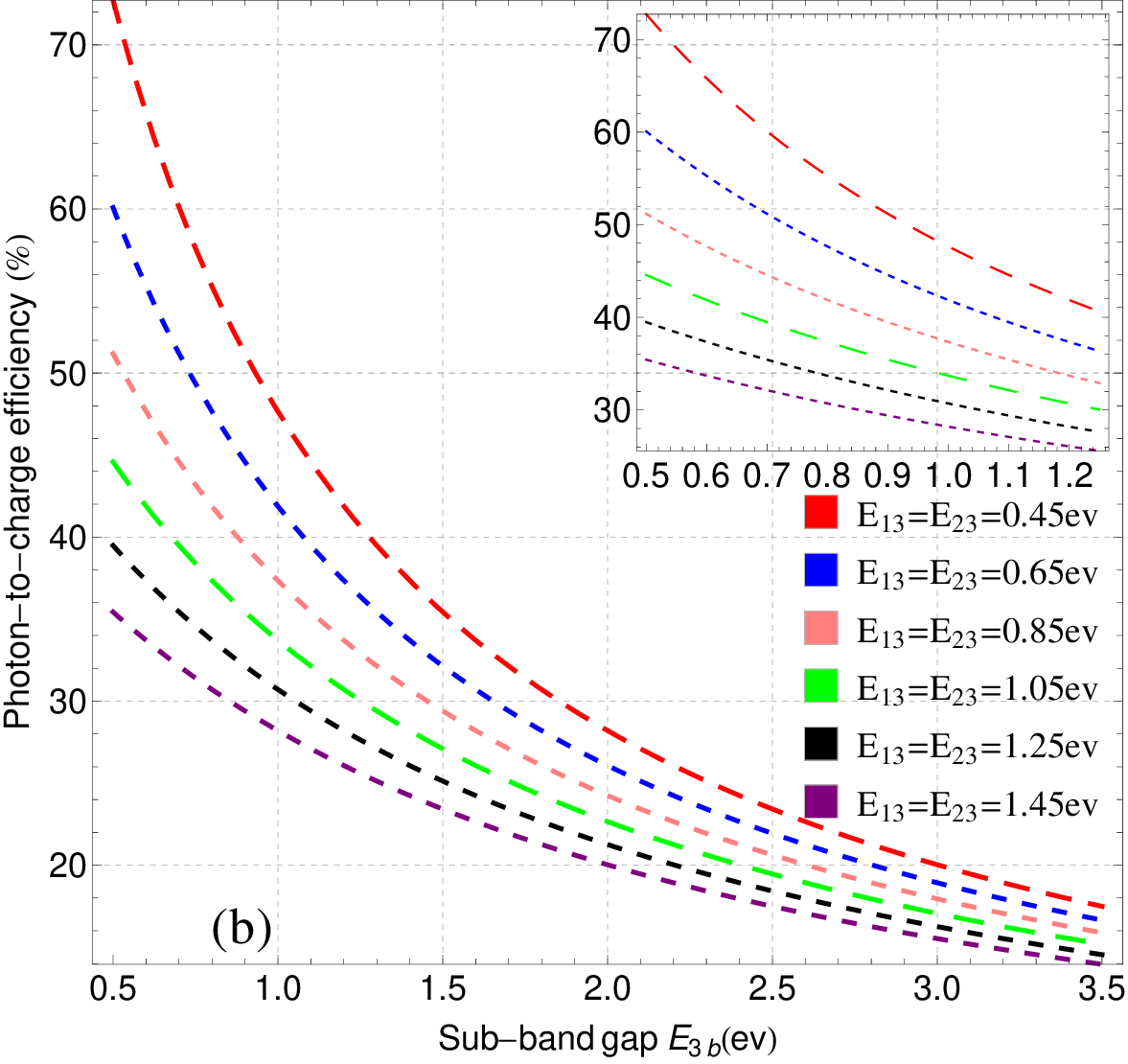}}
\caption{(Color online) Power P and photo-to-charge efficiency generated by the QD-IB photocell versus the lower sub-band gaps(\(E_{3b}\)) for different upper sub-band gaps (\(E_{i3(i=1,2)}\)).}
\label{fig2}
\end{figure}

Figs.2 (a) and (b) show the output power and photo-to-charge efficiency across the load as a function of the lower sub-band gap \(E_{3b}\), and the insert diagrams depict the output power and photo-to-charge efficiency when the lower sub-band gap \(E_{3b}\) absorbs the solar photons within the range of [0.5ev, 1.25ev], whose energy is below the band gaps mentioned in the Shockley and Queisser efficiency\cite{12}. Fig.2 (a) shows an enhancement power as compared to the single-band gap photocell\cite{2,6}, and shows a steep rise for the output power about in the range [0.5ev, 2.5ev] while its growth is slowing down within the range [2.5ev, 3.5ev]. The insert diagram in Fig.2 (a) displays the enhanced power by increasing the lower sub-band gaps(\(E_{3b}\)) within [0.5ev, 1.25ev]. The enhanced power demonstrates that photons with energy below the band gap can contribute to the cell photocurrent on account of the IB. Owing to the IB isolation from the VB with a zero density of states, the carrier relaxation to the VB become difficult and more electrons are stored in the IB\cite{28}. However, the increasing carriers form the delocalized state in the IB and increase their density of states until their wave functions overlap each other and facilitate Mott transition\cite{30}. This process brings out the smooth solid lines within the range [2.5ev, 3.5ev] in Fig.2 (a). What's more, the much higher energy photons within the range [2.5ev, 3.5ev] produce much more thermal phonons which also decreases the growth of the output power. However, the upper sub-band gap \(E_{13}\)=\(E_{23}\) display an opposite evolutionary behavior when it absorbs the solar photons with energies ranging from 0.5 eV to 1.5 eV. And Figs.2 (a) shows the maximum output power with the upper sub-band gaps \(E_{13}\)=\(E_{23}\)=0.45 ev while the minimum output power with the upper sub-band gaps \(E_{13}\)=\(E_{23}\)=1.45 ev. These results indicate that the narrower upper sub-band (\(E_{13}\)=\(E_{23}\)) can easily excite more charge carriers to the CB and produces an enhanced photocurrent. The phenomena is physically caused by the doping IB which adds a ladder for the solar photons below the semiconductor band gap, and much charge carriers stored in the IB can easily transit to the CB with little input energy.

Does the greater output power imply the greater photo-to-charge efficiency \(?\) The photo-to-charge efficiency in Fig.2 (b) displays an opposite evolutionary behavior within the range from 0.5 eV to 3.5 eV. It notes that the photo-to-charge efficiency decreases accompanying with the wider and wider lower sub-band gap \(E_{3b}\). When the the lower sub-band gaps \(E_{3b}\) equals 0.5ev, the photo-to-charge efficiency achieves the maximum peak around 72.4\(\%\) and 36.0\(\%\) with the upper sub-band gap \(E_{13}\)=\(E_{23}\)=0.45ev, 1.45ev, respectively. However, the minimum conversion efficiencies are below 20.0\(\%\) with the lower sub-band gap \(E_{3b}\) =3.5ev when the upper sub-band gaps \(E_{13}\)=\(E_{23}\) are set as 0.45ev and 1.45ev, respectively. And the insert diagram in Fig.2 (b) displays much higher photo-to-charge efficiency than the Shockley and Queisser efficiency\cite{12} when the lower sub-band gap \(E_{3b}\) absorbs photons in the range of [0.5ev, 1.25ev]. The results indicate that the wider lower sub-band gap \(E_{3b}\) can promote the output power but weaken the photo-to-charge efficiency, while the narrower upper sub-band gap \(E_{i3(i=1,2)}\) promotes not only the output power but also the photo-to-charge efficiency. And the main underlying physical significance comes from the absorbed solar photons. The higher energy photons can excite multi-exciton to the IB which leads to much more electron carriers shored in the IB, then the enhanced power is advent eventually. Meanwhile, the absorbed higher photons incur the decreasing ratio between the output energy and the incident solar photons energy, i.e., the photo-to-charge efficiency decreasing with the much higher solar photons.

It is noted that Fig.2 has examined the quantum yield when this QD-IB photocell was manipulated by the upper or lower sub-band gaps with a fixed output voltage, and the sub-band gap layout is the narrow upper sub-band gap accompanying with the wider lower sub-band gap. However, the quantum yield of the QD-IB photocell with an opposite sub-band gap layout is also worthy attention. Fig.3 shows the quantum yields versus the output voltage with a different layout of sub-band gaps, i.e., \(E_{13}\)=\(E_{23}\)=1.43ev and \(E_{3b}\)=0.25ev. Considering the actual ratio has an order of magnitude around 20 between the solar temperature and the ambient temperature circumstance, the solar temperature in this toy model is set as \(T_{s}\)=7ev comparing to the selected ambient temperature values in Fig.3 (And the data for  Fig.3 are in the SI Word), and other parameters are the same to those in Fig.2.

\begin{figure}
\centerline{\includegraphics[width=0.50\columnwidth]{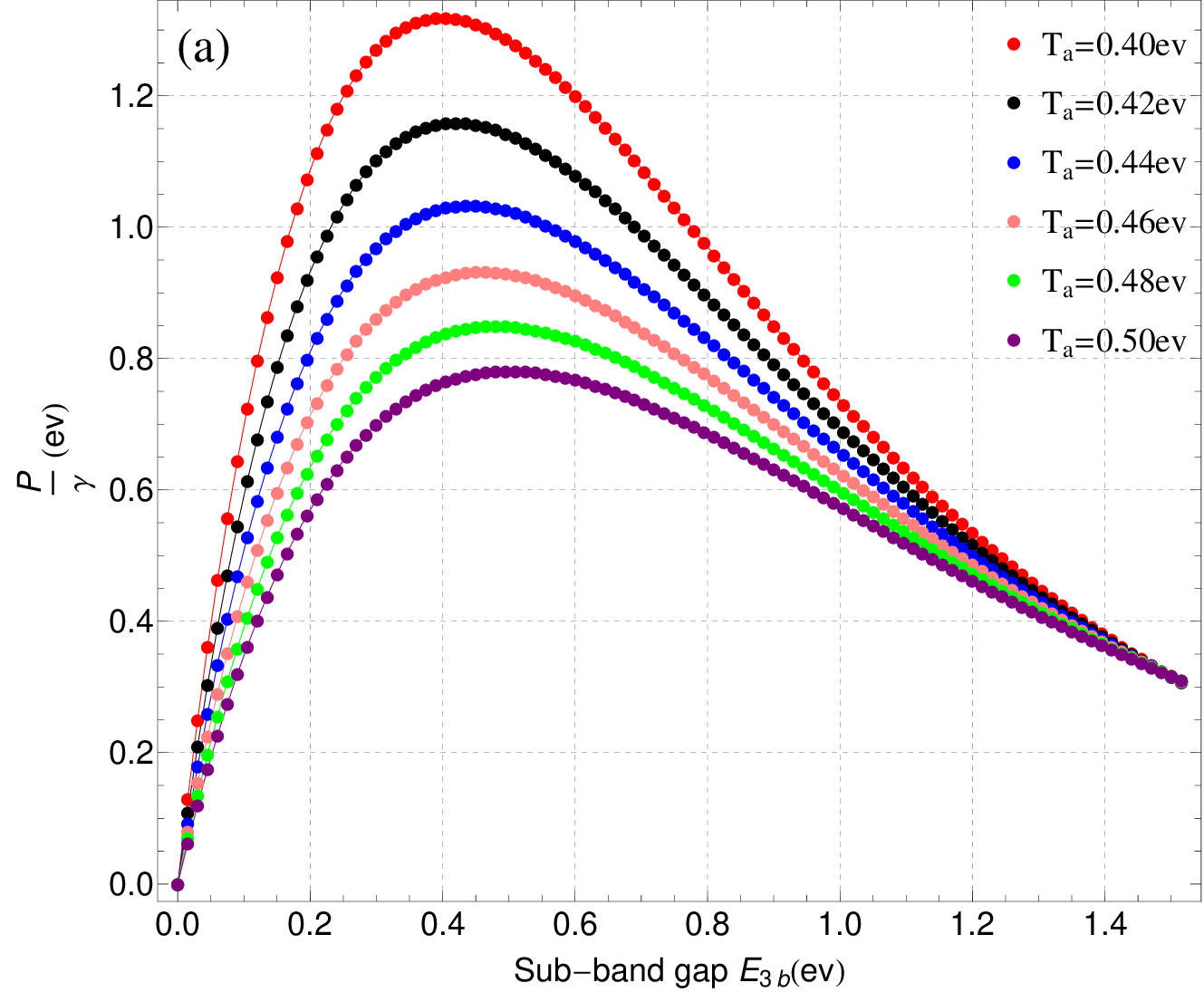}~~~~\includegraphics[width=0.466\columnwidth]{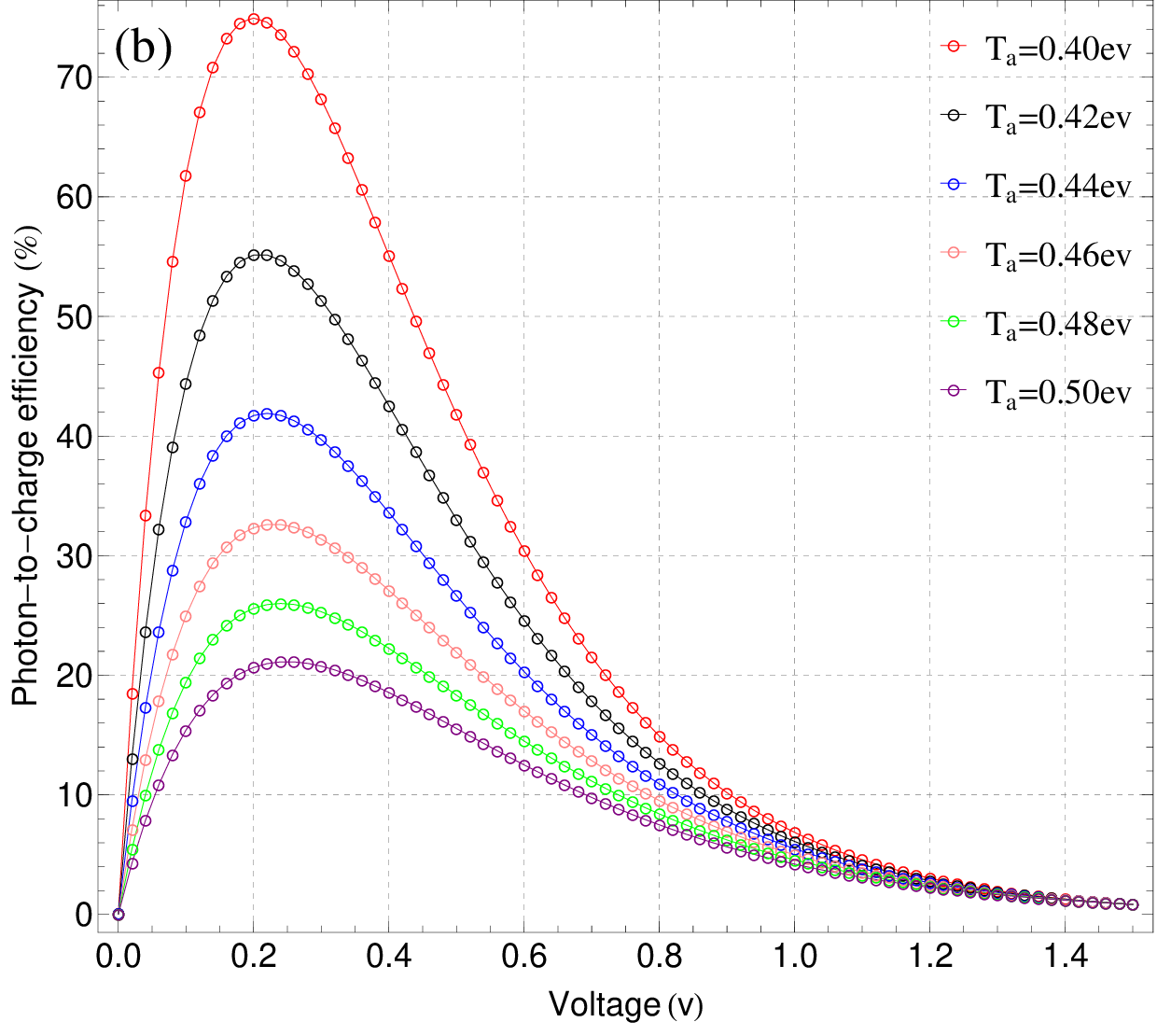}}%
\caption{(Color online) Power P and photo-to-charge efficiency generated by the QD-IB photocell versus the output voltage \(V\) under the different ambient temperatures \(T_{a}\).}
\label{Fig.3}
\end{figure}

Totally different behavior for the quantum yields are advent in Fig.3. Under the specify finite temperature, the output power sharply increases to the peak firstly but then drops to the minimum slowly, and the peak value decreases by an ambient temperature increment being 0.02ev in Fig.3(a). At the ambient temperature \(T_{a}\)=0.40ev, the output power rapidly increases to the peak within the voltage range [0, 0.4v] and slowly declines to the minimum within the voltage range [0.4v, 1.5v], and it outputs the maximum power at the voltage around 0.4v. While at the highest ambient temperature \(T_{a}\)=0.50ev, the curve for the power has the similar distribution pattern except the much smaller peak within the voltage range [0, 1.5v]. Fig.3(a) indicates that the photocurrent can be enhanced by the increasing of the voltage in the range of [0, 0.4v] while the photocurrent decrease when the voltage exceeds 0.4v, and these results show the negative influence of the ambient temperature on the output power. It demonstrates that the higher temperature circumstance can improve the collision probability then decrease the photon-generated carrier transferring across the load, which results in the decreased output power. Although the similar distribution curves for the photo-to-charge efficiency is shown in Fig.3(b), it observably displays the voltages corresponding to peak efficiencies are around 0.2v, as is different from the voltage corresponding to the peak power in Fig.3(a), which demonstrates the asynchrony of the voltages corresponding to the peak power and peak efficiency. What's more, it shows the peak efficiency is significantly affected by the ambient temperature circumstance. The peak photo-to-charge efficiency reaches 74.9\(\%\) at the lowest ambient temperature circumstance \(T_{a}\)=0.40ev, while it declines to 21.21\(\%\) at the maximum temperature circumstance \(T_{a}\)=0.50ev. These indicates the higher ambient temperature circumstance can produce much hot electrons and hot holes, and the increasing thermalization loss destroy the photo-to-charge conversion. Then the efficiencies begin to decline at around 0.2v, which incurs the efficiencies are much smaller at the voltage V=0.4v than those at the voltage V=0.2v.

This QD-IB photocell displays the output power has a prominent enhancement than the single-band gap photocell\cite{2,6} in two different sub-band gap layouts, and the achieved maximum quantum photo-to-charge conversion efficiencies are higher than the limiting efficiency of 47\(\% \) at 1 sun, 63.2\(\% \) at full solar concentration for the IB approach in the former work Ref.\cite{23}. What's more, the maximum quantum photo-to-charge conversion efficiency are also higher than 43\(\% \) at 1 sun, 55\(\% \) for full solar concentration\cite{23} for two-gap tandem solar cell. However, it should be noted that this study has focus only on the quantum yields for the photons below the energy gap
or for ambient operating temperature circumstance, the lost efficiency due to thermalization and the inter-band transition rate and surface recombination rate of carriers were not discussed here. Designing a model of quantum thermalization for photocell and with the consideration of inter-band transition rate and surface recombination rate of carriers will be explicitly discussed in our coming analysis work.

\section{Summary and discussion }
In conclusion, the quantum yields were discussed in this proposed QD-IB photocell. Owing to the IB inserted between the CB and VB, the higher quantum yields were achieved in two different sub-band gap layouts between the upper sub-band gap \(E_{13}\)=\(E_{23}\) and the lower sub-band gap (\(E_{3b}\)). The achieving output power has a prominent enhancement than the single-band gap photocell, and the achieving peak photo-to-charge efficiency reaches to 74.9\(\%\) as compared to the limit efficiency of 63.2\(\%\) via the IB approach in the theoretical solar cell prototype. And the quantum behavior of thermalization, inter-band transition rate and surface recombination rate of carriers in the quantum photocell may be other effective parameters to promote the quantum yields, they are worthy of attention in the QD-IB photocell in the future.

\section*{Acknowledgment}

We gratefully acknowledge support of the National Natural Science Foundation of China ( Grant Nos. 61205205 and 61565008 ),
the General Program of Yunnan Applied Basic Research Project, China ( Grant No. 2016FB009 ).





\bibliographystyle{elsarticle-num}
\bibliography{<your-bib-database>}


\end{document}